\documentstyle[twoside,fleqn,espcrc2,psfig]{article}

\title{Symmetry Breaking, Duality and Fine-Tuning in Hierarchical Spin 
Models
\thanks{Presenting author: Y. Meurice. This work is
      supported in part by the Department of Energy
under Contract No. FG02-91ER40664. }}

\author{J. J. Godina
\address{
Departamento de F\'isica, CINVESTAV-IPN, Ap. Post. 14-740, Mexico, D.F. 07000},
Y. Meurice\address{Department of Physics and Astronomy, University of Iowa, 
Iowa City, Iowa 52242, USA},
S. Niermann$^{\rm b}$ and M. B. Oktay$^{\rm b}$ }

\begin{document}

\begin{abstract}
We discuss three questions related to the critical behavior of 
hierarchical spin models:
1) the hyperscaling relations in the broken symmetry phase; 2) the 
combined use of dual
expansions to calculate non-universal quantities; 3) the fine-tuning issue
in approximately supersymmetric models.

\end{abstract}

\maketitle

\section{Introduction}

The Monte Carlo method has played a major role in the 
understanding of quantum field theory problems for which 
there are no known reliable series expansions. However, the
errors 
associated with this method usually decrease like $t^{-1/2}$
where $t$ is the CPU time used for the calculation. 
This often prohibits obtaining high-accuracy results.
If in the next decades a better knowledge of the fundamental
laws of physics has to rely more and more on precision tests, one 
should complement the Monte Carlo methods with new computational 
tools which emphasize numerical accuracy. 
This line of reasoning is a motivation to use ``hierarchical approximations''
(an example is Wilson's approximate recursion formula \cite{wilson}) 
as a starting point
since the RG transformation can be handled easily. 
Before attacking the hard problem of the improvement 
of such approximations, we would like to show that they
allow spectacular numerical accuracy, namely errors decreasing like
${\rm e}^{-A t^u}$, for some positive constant $A$ of order 1 when $t$
is measured in minutes of CPU time and $0.5\leq u\leq 1$.

For definiteness, we consider 
Dyson's hierarchical model where the total spin in boxes of size
$2^l$ are coupled with a strength $({c\over 4})^l$, with $c$ 
a free parameter set to  
$2^{(D-2)/D}$ in order to approximate
a nearest neighbor scalar model in $D$-dimensions 
(see Ref. \cite{finite} for details).
These interactions are obviously not ultralocal.
Models with Fermi fields \cite{susy} can be constructed similarly
by replacing $D-2$ by $D-1$ in $c$. In addition one needs to specify
a local measure $W_0(\phi)$, for instance of the 
Landau-Ginzburg type ($W_0(\phi) =  e^{-A\phi^2-
B \phi ^4}$) or of the Ising type.
Under a block spin transformation, the local measure changes 
according to
\begin{eqnarray}
&&W_{n+1}(\phi)\propto e^{{\beta \over 2} ({c\over 4})^{n+1} \phi ^2}\nonumber \\
&&\times\int d\phi' W_n({{(\phi -\phi ')}\over 2})
W_n({{(\phi +\phi ')}\over 2}) \ .
\end{eqnarray}
This recursion formula can be reexpressed in Fourier representation as
\begin{equation}
\label{eq:rec}
R_{n+1}(k)\propto exp(-{1\over 2}\beta 
{{\partial ^2} \over 
{\partial k ^2}})(R_{n}({{\sqrt{c}k}\over 2}))^2 \ , 
\end{equation}
where $R_n(k)$ is the Fourier transform of $W_n(\phi)$ with 
an appropriate rescaling. 
It was found \cite{finite} that polynomial
approximations of reasonably small degree 
provide very accurate results in the symmetric phase.
The initial coefficients are the only 
integrals which needs to be calculated, after 
we only perform {\it algebraic} manipulations.
The effects of finite dimensional truncations decay faster than
${\rm e}^{-A_1\sqrt{t}}$ and 
the finite-size effects like ${\rm e}^{-A_2t}$
Direct fits of the susceptibility give a value \cite{gamm} of the 
critical exponent
$\gamma=1.299140730159$. All the digits of this result are confirmed 
by a calculation of the eigenvalues of the linearized 
RG transformation about the non-trivial 
fixed point \cite{wittwer} of Eq. (\ref{eq:rec}).

Unfortunately, is not possible to get rid of the
finite size effects in the broken symmetry phase \cite{gamm} 
by iterating
Eq. (\ref{eq:rec}) enough times because as one moves away from 
the fixed point, 
rapid oscillations appear. Namely,
$R_n(k)\simeq {\rm cos}(m c^{n/2}k)$, where $m$ is approximately the 
magnetization at zero external field. 
It is easy to see from Eq. (\ref{eq:rec}), that for $n$ large enough, the 
polynomial approximation breaks down as the argument of the exponential 
becomes too large. It is nevertheless 
possible to take advantage of the short 
number of iterations for which the low temperature scaling is observed to 
obtain reliable extrapolations, first to infinite volume 
and non-zero external field and then to zero
external field. Proceeding this way, one can calculate the connected 
$q$-point functions at zero momentum 
$G_q^c(0)$,
for $q=1$, 2 and 3 and for various values of $\beta$.
The scaling and hyperscaling relations together with $\eta=0$ imply
\begin{equation}
G_q^c(0)\propto(\beta-\beta_c)^{(\gamma/2)(D(1-q/2)-q)} \ ,
\end{equation}
where $\gamma$ is the value calculated in the symmetric phase.
Our numerical estimate of the exponents \cite{lt} agree
with the predicted values of the exponents with an accuracy
of $10^{-3}$, which is respectable when compared to results obtained 
with conventional methods, but large compared to the $10^{-12}$ errors 
obtained in the symmetric phase. 
In order to improve the accuracy of these results, one could increase 
the degree of the polynomial 
used for the approximation, however a short calculation
shows that with this procedure, the errors only decrease like $1/t$.
A more appealing possibility consists in factorizing $R_n(k)$ into two part:
one rapidly oscillating and which requires an exact treatment and another 
slowly varying which can be approximated by polynomials. This second 
possibility is presently under investigation.

We now discuss the possibility of replacing the numerical evaluation (which
has to be repeated for each choice of the input parameters) by a suitable 
series expansion (which could be done once for all). Our goal is to
obtain an analytical formula for the non-universal quantities 
$(A_0,\  A_1\ldots)$ appearing in
the parametrization of the susceptibility:
\begin{equation}
\chi\simeq (\beta _c -\beta )^{-\gamma } (A_0 + A_1 (\beta _c -\beta)^{
\Delta }+\ldots )\ .
\label{eq:param}
\end{equation}
The linearization method used to calculate the universal exponents 
$\gamma$ and $\Delta$ does not provide a way to calculate
the non-universal quantities $(A_0,\  A_1\ldots)$. Indeed, their values 
``build up''
during the crossover between the unstable IR fixed point and the 
high-temperature (HT) fixed point. A proper description of this crossover
requires non-linear expansions about $both$ fixed points.
This problem can be solved completely \cite{dual} for the simplified
recursion relation for the magnetic susceptibility 
\begin{equation}
\chi_{n+1}=\chi_n +(\beta/ 4) ({c/2})^{n+1}\chi_n^2\ ,
\label{eq:sus}
\end{equation}
which was used in Ref. \cite{finite}
to estimate the finite volume effects.
In this simplified model, one can construct non-linear functions $y$
and $\tilde{y}$ which transform covariantly (multiplicatively)
under a RG transformation. 
These two functions are expressed as an expansion about the IR 
and HT fixed point respectively.
It turns out that there exists \cite{dual} a
duality
relation between the series expansions of the two functions. 
It would be very interesting to see if similar methods could be applied 
for non-linear sigma models or gauge theories.
The quantity $A_0$ is a function of $(\beta_c-\beta)$ with a discrete scale 
invariance and it can be expanded in Fourier modes as in the original 
model \cite{high}.
For practical purposes, the contribution of the non-zero modes is 
exponentially suppressed and in very good approximation:
\begin{equation}
A_0={1\over{ \ln \lambda}}\int_{z_{a}}^{\lambda z_{a}}(dz/z)
z^{\gamma} \tilde{y}[1-y^{-1}(z)]\ ,
\end{equation}
where $\lambda$ is the eigenvalue associated with the unstable direction of the
IR fixed point.
The lower value $z_{a}$ of the integration interval is arbitrary and we 
can choose it
at our convenience.
We have compared 
the approximate values $A_0(m,\tilde{m})$ 
obtained from expansions 
with $m$ terms for $y^{-1}$ and $\tilde{m}$ terms for $\tilde{y}$
with an accurate value of $A_0$ and found that the errors go approximately 
like ${\rm exp}(-K_1(m+\tilde{m})+K_2(m-\tilde{m})^2)$.
This implies that for $m+\tilde{m}$ fixed, it is 
very advantageous to pick the ``self-dual'' option $m\simeq\tilde{m}$.

We now discuss the fine-tuning question. 
The fact that the bare mass of a scalar field theory requires a
fine-tuning in order to keep the renormalized mass small in cut-off units
is usually regarded as an argument against fundamental scalars.
A possible resolution of this feature consists in adding 
degrees of freedom in such a way that the
quantum fluctuations cancel, making small scalar masses a more 
natural outcome. 
In a recent preprint \cite{susy}, we presented 
two models with an approximate supersymmetry and 
which can be solved
non-perturbatively with the numerical methods discussed above.
The bosonic part of these models
is Dyson's hierarchical model.
The free action for $N$ massless scalar fields 
$\phi_x^{(i)}$ reads
\begin{equation}
S_B^{free}={1\over 2}\sum_{x,y,i}\phi_x^{(i)}D^2_{xy}\phi_y^{(i)}\ ,
\end{equation}
where $x$ and $y$ run over the sites and $i$ from 1 to $N$. 
The action for free massless fermions reads
\begin{equation}
S_F^{free}=\sum_{x,y,i}\bar{\psi}_x^{(i)}D_{xy}\psi_y^{(i)}\ ,
\end{equation}
where the $\psi_x^{(i)}$ and  $\bar{\psi}_x^{(i)}$ are Grassmann numbers.
The explicit form of $D_{xy}$, its square $D^2_{xy}$ and the supersymmetry
of the free action are given in \cite{susy}.

The Grassmann nature of the fermionic fields restricts severely the type
of interactions allowed. For instance, for one flavor the most
general local measure (without any supersymmetry considerations) is 
\begin{equation}
{\cal{W}}(\phi,\psi,\bar{\psi})=W(\phi)+\psi\bar{\psi}A(\phi)
\end{equation}
A more interesting local measure with two-flavors and a R-symmetry is
given in \cite{susy}. The RG transformation can be expressed as a finite number
of convolutions which can be calculated accurately in Fourier
transform with polynomial approximations as in the bosonic case 
(see \cite{susy} for explicit formulas). Some numerical results are 
summarized in Fig. 1 where the renormalized mass is plotted as function
of the bare mass (both in cut-off units).
\begin{figure}[htb]
\centerline{\psfig{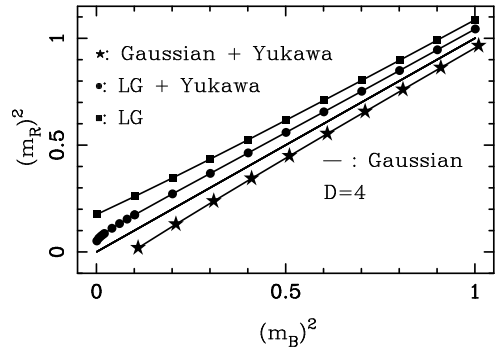}}
\caption{The renormalized mass ($m_R$) as a function of the bare mass ($m_B$)
for various models.}
\end{figure}
One sees that in the purely bosonic interacting model (LG), 
the quantum corrections 
to $m_R^2$ are positive while for a Gaussian model coupled to fermions
with a Yukawa coupling, the corrections are negative. 
For visual reference, the bosonic Gaussian result, 
where the two quantities are 
obviously equal, is also plotted (straight line).
We would like to know if by combining the two interactions in an appropriate
way, it possible to cancel the two corrections. 
At the one-loop level, a simple relation ($g_y^2=8\lambda _4$) guarantees that
$m_R$ goes to zero when $m_B^2$ goes to zero. However, we 
found numerically that for such a choice, 
$m_R^2\simeq 0.044$ (in cutoff units) 
when $m_B^2$ goes to zero.
It is possible to fine-tune $g_y$ in order to get $m_R=0$ and 
the exact critical value of $g_y$ is about 50 percent larger
than the perturbative one. We are presently trying to build models
with better naturalness properties.

\end{document}